\documentclass[aps,twocolumn,showkeys,superscriptaddress,amsmath,amssymb]{revtex4}

\usepackage{graphicx}
\usepackage{dcolumn}
\usepackage{bm}
\usepackage{color}

\begin{document}

\title{LiH as a Li$^{+}$ and H$^{-}$ ion provider}

\author{Khang Hoang}
\affiliation{Materials Department, University of California, Santa Barbara, California 93106, USA}%
\affiliation{Computational Materials Science Center, George Mason University, Fairfax, Virginia 22030, USA}%
\affiliation{Center for Computationally Assisted Science and Technology, North Dakota State University, Fargo, North Dakota 58108, USA.}%
\author{Chris G. Van de Walle}
\affiliation{Materials Department, University of California, Santa Barbara, California 93106, USA}%


\begin{abstract}

We present a first-principles study of the formation and migration of native defects in lithium hydride (LiH), a material of interest in hydrogen storage and lithium-ion batteries. We find that the negatively charged lithium vacancy and positively charged hydrogen vacancy are the dominant defects in LiH with a formation energy of about 0.85 eV. With this low energy, LiH can be a good provider of Li$^{+}$ and/or H$^{-}$ ions. Based on our results, we discuss the role of LiH in its reactions with another reactant which are important in hydrogen storage and lithium-ion batteries, and explain the ionic conductivity and hydrogen surface loss as observed in experiments.

\end{abstract}

\keywords{lithium hydride, hydrogen storage, lithium-ion batteries, defects, kinetics, first principles}
\maketitle

\section{\label{sec:intro}Introduction}

Lithium hydride (LiH) is in principle of interest for hydrogen storage~\cite{ANIE:ANIE200806293} because of its very large hydrogen content. However, with a decomposition temperature of 720$^\circ$C \cite{Grochala}, it is not very useful for practical applications. The use of LiH in combination with other complex metal hydrides, on the other hand, has been shown to lower the dehydrogenation temperature and improve hydrogen kinetics~\cite{chenNATURE,Leng2004}. For instance, the lithium amide/imide reaction has been reported to reversibly store and release about 6.5 wt\% hydrogen during absorption at 20 bar and desorption at 0.04 bar at temperatures below 300$^\circ$C~\cite{chenNATURE}:
\begin{equation}\label{eq:amide}
\mathrm{LiNH}_{2} + \mathrm{LiH} \leftrightarrow \mathrm{Li}_{2}\mathrm{NH} + \mathrm{H}_{2}.
\end{equation}
LiH is also present in lithium-ion battery electrodes that are based on conversion reactions \cite{Oumellal}, e.g.,
\begin{equation}\label{eq:li-ion}
M\mathrm{H}_{x} + x\mathrm{Li}^{+} + xe^{-} \leftrightarrow M + x\mathrm{LiH},
\end{equation}
where $M$ is a metal or an intermetallic compound. In both cases, the hydrogen storage or electrochemical performance ultimately depends on the reaction kinetics, which is controlled by mass transport mediated by native defects in the reactant materials. Other experimental observations include the production of H$^{-}$ ions by heating LiH in vacuum~\cite{siefken:3394}, and loss of hydrogen from LiH surfaces at room temperature, causing the formation of a lithium metal phase~\cite{Powell1974345}. Regarding the ionic conductivity, LiH was reported to have an activation energy of 1.70$\pm$0.10 eV and a cation vacancy migration barrier of 0.54$\pm$0.02 eV~\cite{Varotsos,Ikeya}.

A detailed understanding of these reactions and phenomena is still lacking, largely due to the absence of information about the formation and migration of native point defects in LiH. Native defects such as Schottky pairs, anion and cation Frenkel pairs, and antisite pairs were studied by Pandey and Stoneham using an interatomic potential model~\cite{stoneham}. To our knowledge, there has not been a first-principles study of the defects, except for the preliminary formation energy value of the negatively charged lithium vacancy $V_{\rm Li}^{-}$ and the positively charged hydrogen vacancy $V_{\rm H}^{+}$ reported in our previous work~\cite{hoang_prb2012}. Here we apply modern first-principles techniques and present a more comprehensive study of native defects in LiH based on density-functional theory (DFT). Besides the neutral defect pairs, we study individual lithium- and hydrogen-related defects such as vacancies, interstitials, and antisite defects, and investigate the dependence of their formation energy (and hence concentration) on the Fermi level and atomic abundances. We find that $V_{\rm Li}^{-}$ and $V_{\rm H}^{+}$ are the dominant defects in LiH, both with a relatively low formation energy. Our results provide insights into the role of LiH in reactions such as Eqs.~(\ref{eq:amide}) and (\ref{eq:li-ion}), and explain the ionic conductivity and hydrogen surface loss as observed in experiments.

\section{\label{sec:metho}Methods}

Our total-energy calculations are based on DFT within the Perdew-Burke-Ernzerhof (PBE) version~\cite{GGA} of the generalized-gradient approximation (GGA) and the projector augmented wave method~\cite{PAW1,PAW2}, as implemented in the VASP code~\cite{VASP1,VASP2,VASP3}. The Heyd-Scuseria-Ernzerhof hybrid functional (HSE06)~\cite{heyd:8207,paier:154709} was also used in some bulk and defect calculations. Calculations for bulk LiH were performed using a 21$\times$21$\times$21 Monkhorst-Pack $\mathbf{k}$-point mesh~\cite{monkhorst-pack}. For calculations of native defects, we used a cubic (3$\times$3$\times$3) 216-atom supercell with a 2$\times$2$\times$2 $\mathbf{k}$-point mesh and a plane-wave basis-set cutoff of 400 eV. The lattice parameters were fixed to the calculated bulk values, but all the internal coordinates were fully relaxed. Convergence with respect to self-consistent iterations was assumed when the total energy difference between cycles was less than 10$^{-4}$ eV and the residual forces were smaller than 0.01 eV/{\AA}. Migration barriers of selected defects were studied using the climbing-image nudged elastic-band (NEB) method~\cite{ci-neb}.

Native defects in LiH can be characterized by their formation energies, defined as~\cite{hoang_prb2012,wilsonshort09,walle:3851}
\begin{eqnarray}\label{eq:eform}
\nonumber
E^f({\mathrm{X}}^q)=E_{\mathrm{tot}}({\mathrm{X}}^q)&-&E_{\mathrm{tot}}({\mathrm{bulk}})-\sum_{i}{n_i\mu_i} \\
&+&q(E_{\mathrm{v}}+\mu_{e})+ \Delta^q ,
\end{eqnarray}
where $E_{\mathrm{tot}}(\mathrm{X}^{q})$ and $E_{\mathrm{tot}}(\mathrm{bulk})$ are, respectively, the total energies of a supercell containing the defect X in charge state $q$ and of a supercell of the perfect bulk material; $\mu_{i}$ is the atomic chemical potential of species $i$ (referenced to bulk Li metal and H$_{2}$ molecules at 0 K), and $n_{i}$ denotes the number of atoms of species $i$ that have been added ($n_{i}$$>$0) or removed ($n_{i}$$<$0) to form the defect. $\mu_{e}$ is the electron chemical potential, i.e., the Fermi level, referenced to the valence-band maximum in the bulk ($E_{\mathrm{v}}$). $\Delta^q$ is the correction term to align the electrostatic potentials of the bulk and defect supercells and to account for finite-cell-size effects on the total energies of charged defects \cite{walle:3851,freysoldt,Freysoldt11}.

The quantities in Eq.~(\ref{eq:eform}) can be computed from the total-energy calculations. The atomic chemical potentials $\mu_{i}$ are variable, but subject to thermodynamic constraints~\cite{walle:3851,wilsonshort09,hoang_prb2012}, such as the condition for stability of LiH:
\begin{equation}\label{eq:stability}
\mu_{\rm Li} + \mu_{\rm H} = \Delta H_{f}(\rm {LiH}),
\end{equation}
where $\Delta H_{f}$ is the formation enthalpy, which is $-$0.825 eV (at 0 K) in our DFT-GGA calculations. In this equation, we have implicitly referenced $\mu_{\rm Li}$ and $\mu_{\rm H}$ to the values for the reference phases: bulk Li and 1/2 an H$_2$ molecule.

In thermal equilibrium, the concentration of the defect X at temperature $T$ can be obtained via the relation~\cite{walle:3851}
\begin{equation}\label{eq:concen}
c(\mathrm{X})=N_{\mathrm{sites}}N_{\mathrm{config}}\mathrm{exp}[-E^{f}(\mathrm{X})/k_BT],
\end{equation}
where $N_{\mathrm{sites}}$ is the number of high-symmetry sites in the lattice per unit volume on which the defect can be incorporated, and $N_{\mathrm{config}}$ is the number of equivalent configurations (per site). Clearly, defects with low formation energies will easily form and occur in high concentrations.

\section{\label{sec:results}Results and Discussion}

LiH crystallizes in the rocksalt structure with an experimental lattice constant of 4.09 {\AA}~\cite{stephens:177}. The calculated lattice constant is 3.99 {\AA} (volume of 15.9 {\AA}$^{3}$ per formula unit) in both GGA and HSE06, 2.5\% smaller than the experimental value and likely due to the neglect of lattice vibrations in our calculations~\cite{martins}. The calculated volume of LiH is actually smaller than that of face-centered cubic Li metal (20.3 {\AA}$^{3}$/f.u.): as hydrogen is introduced into the Li metal, the lattice spacing decreases due to the strong interaction between Li$^{+}$ and H$^{-}$ ions. 

The implementation of finite-cell-size corrections in the Freysoldt scheme \cite{freysoldt,Freysoldt11} requires values for the static dielectric constant. We find that the electronic contribution to the static dielectric constant of LiH is 4.32 in GGA and 3.64 in HSE06 based on the real part of the dielectric function $\epsilon_{1}$ for $\omega\rightarrow0$. The ionic contribution is 10.65 in GGA, obtained from density functional perturbation theory~\cite{dielectricmethod}. Since the ionic contribution only depends on the Born effective charges and the vibrational modes, which are usually well described in GGA~\cite{CZTS}, this value can be used for both GGA and HSE06 calculations. The calculated total static dielectric constant is thus 14.97 in GGA and 14.29 in HSE06. For comparison, the experimental dielectric constant is 12.9$\pm$0.5~\cite{dielectric}.

\begin{figure}
\vspace{-0.2in}
\begin{center}
\hspace*{-0.10in}
\includegraphics[width=3.6in]{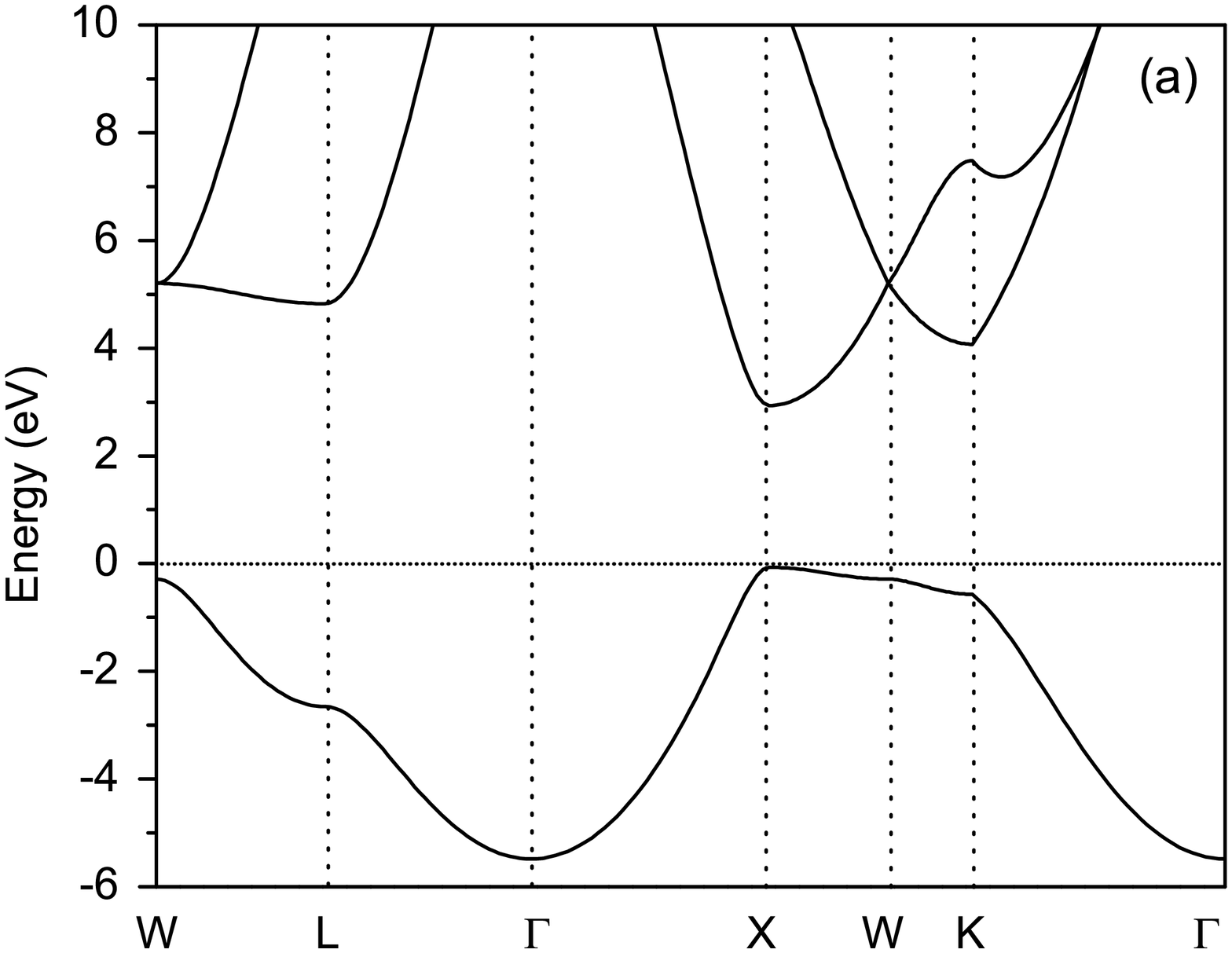} 
\end{center}
\vspace{-0.90in}
\begin{center}
\hspace*{-0.10in}
\includegraphics[width=3.6in]{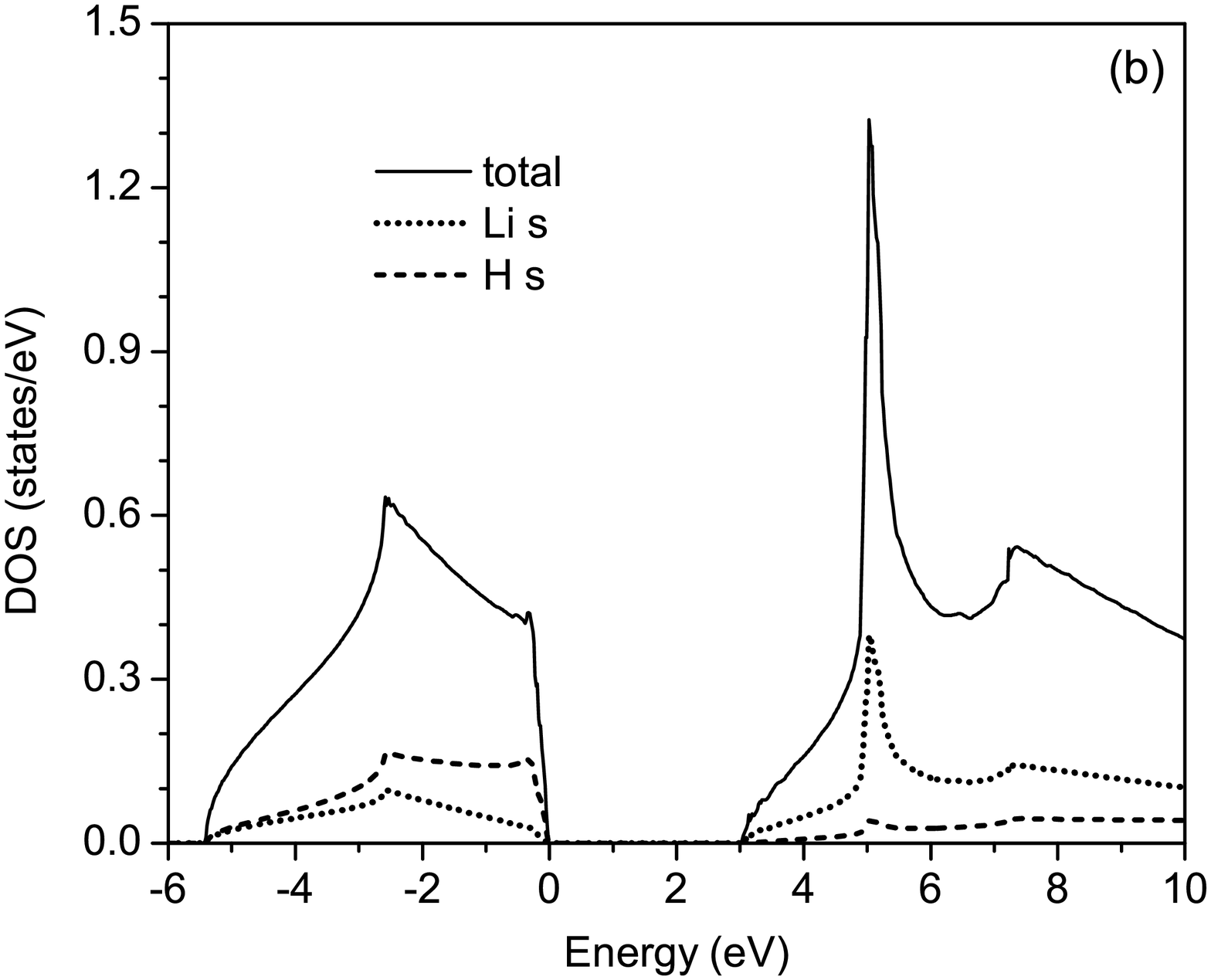} 
\end{center}
\vspace{-0.25in}
\caption{(a) Band structure and (b) total and projected electronic density of states (DOS) of LiH, calculated with DFT-GGA. The zero of energy is set to the highest occupied state.}\label{bulk}
\end{figure}

Figure~\ref{bulk} shows the band structure and the total and projected electronic density of states (DOS) of LiH. The valence band consists of the bonding state of Li and H $s$ states, whereas the conduction band consists of the antibonding state. The valence-band width is 5.41 eV (or 6.09 eV in HSE06). The calculated band gap is 3.02 eV (4.02 eV in HSE06), a direct gap at the $X$ point in the face-centered cubic Brillouin zone. The experimental band gap is 4.99 eV at 4.2 K~\cite{plekhanov}. The projected DOS [Fig.~\ref{bulk}(b)] shows that the states near the valence-band maximum have mainly hydrogen character, while those near the conduction-band minimum have predominantly lithium character, consistent with the notion that lithium acts as the cation (i.e., Li$^{+}$) and hydrogen as the anion (i.e., H$^{-}$). The calculated formation enthalpy of LiH at 0 K is $-$0.825 eV (or $-$0.879 eV in HSE06); for comparison, the experimental value at 300 K is $-$0.94 eV~\cite{NIST-JANAF}. Our GGA results for the bulk properties are in good agreement with those reported by other groups~\cite{vansetten_prb_2007,PhysRevB.75.014101}.

\begin{figure}
\vspace{-0.20in}
\begin{center}
\hspace*{-0.10in}
\includegraphics[width=3.6in]{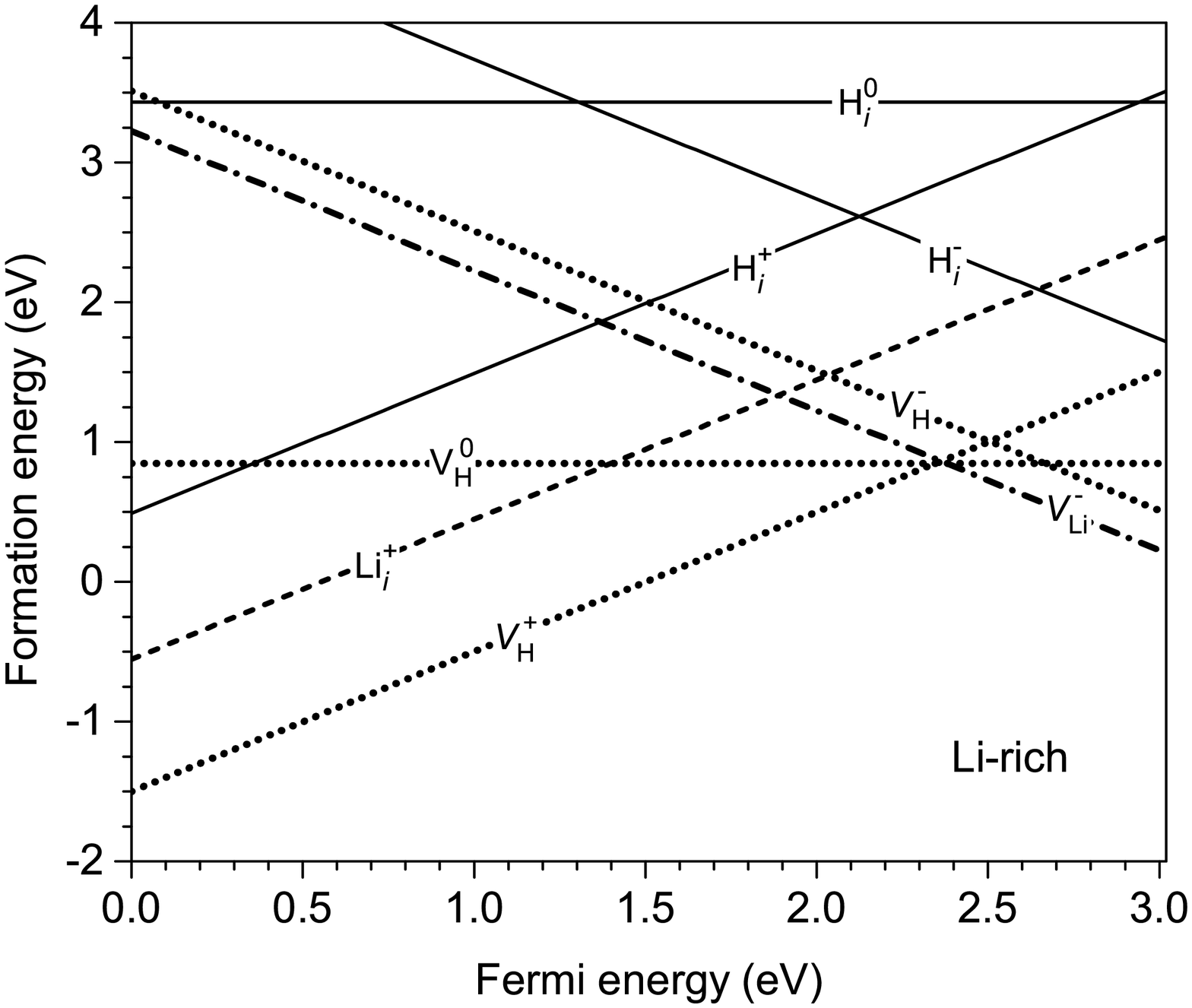} 
\end{center}
\vspace{-0.7in}
\begin{center}
\hspace*{-0.10in}
\includegraphics[width=3.6in]{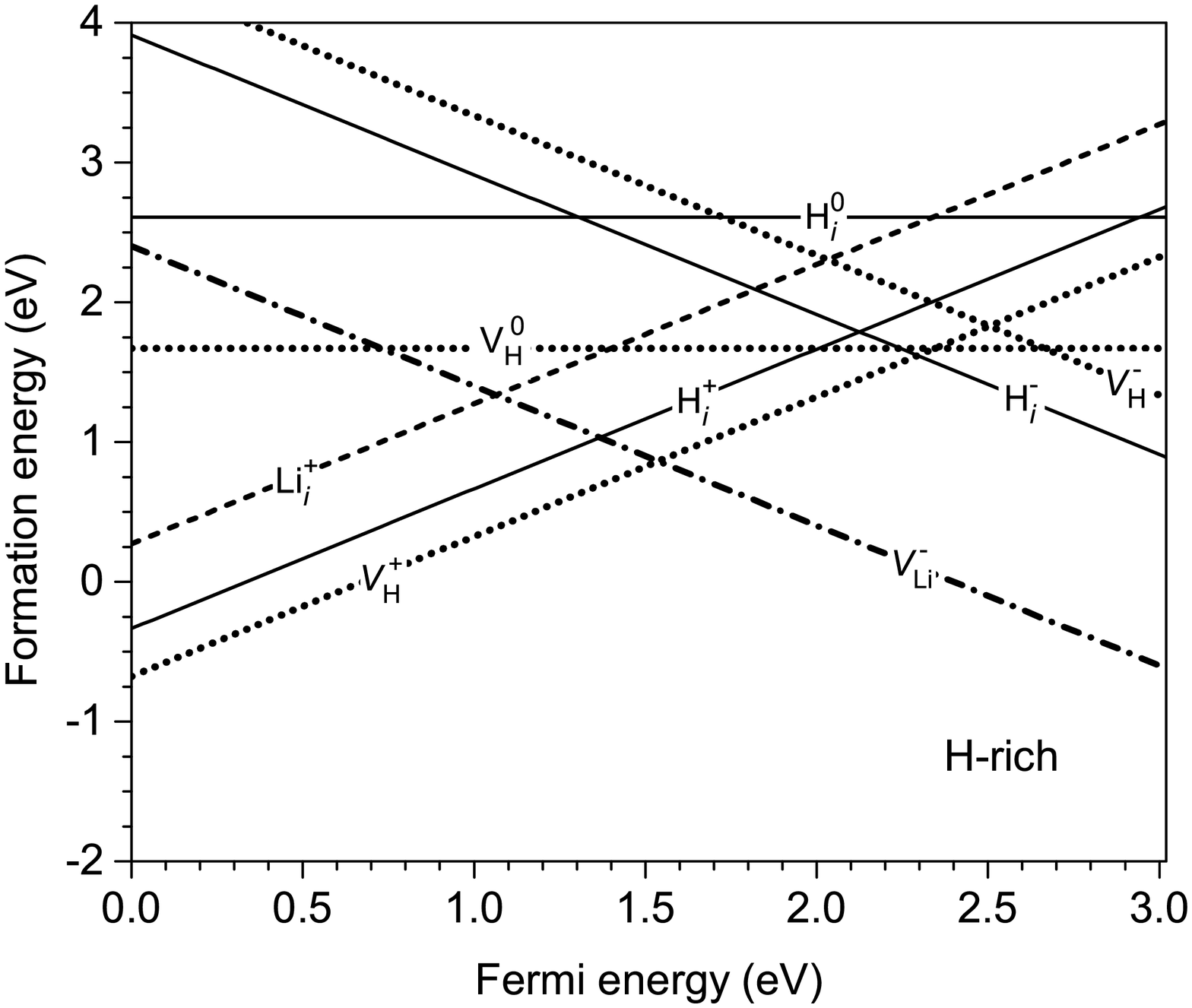} 
\end{center}
\vspace{-0.28in}
\caption{Calculated formation energies of native defects in LiH under Li-rich and H-rich conditions, plotted as a function of Fermi level with respect to the valence-band maximum.}\label{FE}
\end{figure}

We investigated lithium- and hydrogen-related single point defects in LiH such as vacancies, interstitials, and antisite defects in all the possible charge states and their defect complexes. Figure~\ref{FE} shows the calculated formation energies of the native point defects obtained in GGA calculations. The ``Li-rich'' condition corresponds to equilibrium between LiH and the Li metal phase (at 0 K), which gives $\mu_{\rm Li}$=0 eV and $\mu_{\rm H}$=$-$0.825 eV (the calculated formation enthalpy of LiH), whereas the ``H-rich'' condition corresponds to equilibrium between LiH and H$_{2}$ (at 0 K), which gives $\mu_{\rm Li}$=$-$0.825 and $\mu_{\rm H}$=0 eV.

For $V_{\rm H}^{+}$, which is created by removing an H$^{-}$ ion from the LiH supercell, we observe outward relaxations of the neighboring Li atoms by 0.20 {\AA}; whereas for $V_{\rm H}^{0}$ the outward relaxation is only 0.07 {\AA}. The creation of $V_{\rm H}^{-}$ results in inward relaxations of the neighboring Li atoms by 0.03 {\AA}. In the case of H$_{i}^0$ and H$_{i}^-$, the extra H atom or ion stays near the tetrahedral site of the rocksalt structure with an average distance to the four neighboring Li atoms of 1.76 {\AA} (H$_{i}^0$) or 1.67 {\AA} (H$_{i}^-$). For H$_{i}^+$, the extra H$^{+}$ ion combines with a host H$^{-}$ ion and forms a dumbbell that has its center at the octahedral site and an H$-$H distance of 0.76 {\AA}. For comparison, the calculated H$-$H bond in an H$_{2}$ molecule is 0.75 {\AA} in both GGA and HSE06 calculations. The binding energy of an H$_{2}$ molecule with respect to spin-polarized H atoms is $-$4.53 eV and $-$4.52 eV in GGA and HSE06, respectively. Significant lattice distortions occur in the neighborhood of these interstitials. We also investigated interstitial H$_{2}$ molecules but found them to be unstable as molecules inside LiH.

Our studies of migration focused on the most relevant point defects (see below) and produced energy barriers of 0.26, 0.15, 0.51, 0.84, and 1.07 eV for H$_{i}^+$, H$_{i}^-$, $V_{\rm H}^{+}$, $V_{\rm H}^{0}$, and $V_{\rm H}^{-}$, respectively, using the NEB method~\cite{ci-neb}. The migration of H$_{i}^-$, for instance, involves moving the H$^-$ ion from one tetrahedral site to another, whereas the migration of $V_{\rm H}^{+}$ involves moving a neighboring H$^{-}$ ion to the vacancy. The migration barrier values for H$_{i}^+$, H$_{i}^-$, and $V_{\rm H}^{+}$ indicate that these defects, once formed, are mobile below room temperature.

For $V_{\rm Li}^{0}$, $V_{\rm Li}^{+}$, or $V_{\rm Li}^{-}$, outward relaxations of the neighboring H atoms occur, by about 0.15 {\AA}. In the case of Li$_{i}^0$, Li$_{i}^+$, and Li$_{i}^-$, the extra Li stays at the tetrahedral site with a distance to the four neighboring H$^{-}$ of 1.72 {\AA}. Li$_{i}^0$ and Li$_{i}^-$ are not included in Fig.~\ref{FE} because they have much higher formation energies than other native defects and are thus unlikely to play a role in LiH. Regarding migration, we find an energy barrier for migration of Li$_{i}^+$ of 0.22 eV, and for $V_{\rm Li}^{-}$ the barrier is 0.51 eV. Our results for the migration of $V_{\rm Li}^{-}$ and $V_{\rm H}^{+}$ are in qualitative agreement with those reported in Ref.~\cite{stoneham}, where the barriers for cation and anion vacancies were found to be 0.42 and 0.40 eV, respectively.

The antisite defects Li$_{\rm H}$ (Li replacing H) and H$_{\rm Li}$ (H replacing Li) were also investigated but not included in Fig.~\ref{FE} because of their high formation energies (2.88 eV or higher, see more below). We find significant lattice distortions associated with these antisite defects.

\begin{table}
\caption{Calculated formation energies ($E^{f}$), migration barriers ($E_{m}$), activation energies ($E_{a}$=$E^{f}$+$E_{m}$), and binding energy ($E_{b}$), all in eV, for native defects in LiH under Li-rich and H-rich conditions. For Schottky, Frenkel, and antisite pairs, the formation energy is defined for an atomic configuration in which the two defects in a pair are at the closest distance consistent with the pair being structurally stable, and thus includes a binding energy, defined as the energy difference between the sum of the formation energies of the constituents and the formation energy of the pair. Migration barriers denoted by an asterisk ($^{\ast}$) are estimated by taking the higher of the migration barriers of the constituents.}\label{tab}
\begin{center}
\begin{ruledtabular}
\begin{tabular}{lcclccc}
Defect&\multicolumn{2}{c}{$E^{f}$}&$E_m$&\multicolumn{2}{c}{$E_{a}$}&$E_{b}$\\
&Li-rich&H-rich&&Li-rich&H-rich& \\
\colrule
H$_{i}^+$ &2.85&1.20&0.26&3.11&1.46	\\
H$_{i}^-$ &2.37&2.37&0.15&2.52&2.52\\
$V_{\rm H}^{0}$ &0.85&1.67&0.84&1.69&2.51\\
$V_{\rm H}^{+}$ &0.86&0.86&0.51&1.37&1.37\\
$V_{\rm H}^{-}$ &1.15&2.80&1.07&2.22&3.87\\
Li$_{i}^+$ &1.81&1.81&0.22&2.03&2.03\\
$V_{\rm Li}^{-}$ &0.86&0.86&0.51&1.37&1.37\\
($V_{\rm Li}^{-}$,$V_{\rm H}^{+}$) &1.15&1.15&0.51$^{\ast}$&1.66&1.66&0.58\\
(H$_i^{-}$,$V_{\rm H}^{+}$) &2.82&2.82&0.51$^{\ast}$&3.33&3.33&0.42 \\
(Li$_i^{+}$,$V_{\rm Li}^{-}$) &2.36&2.36&0.51$^{\ast}$&2.87&2.87&0.31 \\
(Li$_{\rm H}^{2+}$,H$_{\rm Li}^{2-}$) &3.05&3.05&&&&3.96\\
\end{tabular}
\end{ruledtabular}
\end{center}
\end{table}

Figure~\ref{FE} shows that under Li-rich and H-rich conditions and any conditions in between these two limits, $V_{\rm Li}^{-}$ and $V_{\rm H}^{+}$ have the lowest formation energies. These vacancies are thus the dominant defects in LiH, i.e., they will occur in the highest concentrations, see Eq.~(\ref{eq:concen}). Other defects are less likely to form because they have higher formation energies. In the absence of extrinsic charged impurities, the Fermi level of the system is at the position $\mu_e$=$\mu_e^{\rm int}$ where the formation energies of the intrinsic defects $V_{\rm Li}^{-}$ and $V_{\rm H}^{+}$ are equal~\cite{wilsonshort09,hoang_prb2012}. Table~\ref{tab} lists the calculated formation energies at $\mu_e^{\rm int}$ as well as migration barriers of selected defects and complexes.

In addition to isolated point defects, we have also investigated neutral defect complexes such as the ($V_{\rm Li}^{-}$,$V_{\rm H}^{+}$) Schottky pair, (H$_i^{-}$,$V_{\rm H}^{+}$) and (Li$_i^{+}$,$V_{\rm Li}^{-}$) Frenkel pairs, and (Li$_{\rm H}^{2+}$,H$_{\rm Li}^{2-}$) antisite pair. We define the formation energy of such pairs to be for an atomic configuration in which the two defects in a pair are at the closest distance consistent with the pair being structurally stable. This implies that our pair formation energy includes a contribution from binding energy between the constituents, defined as the energy difference between the sum of the formation energies of the constituents and the formation energy of the pair. The calculated formation energies and binding energies with respect to their constituents are included in Table~\ref{tab}. A lower bound on the migration barrier of the complexes can be estimated by taking the higher of the migration barriers of the constituents~\cite{wilsonshort09} and is also listed in Table~\ref{tab}. The binding energy of the antisite pair is large (3.96 eV) because when Li$_{\rm H}^{2+}$ and H$_{\rm Li}^{2-}$ form the nearest-neighbor pair the lattice distortion effects cancel out thus lowering the energy of the pair. The calculated formation energies of the dissociated anion Frenkel, cation Frenkel, Schottky, and antisite pairs (i.e., when the two point defects in a pair are infinitely separated) are 3.23, 2.67, 1.72, and 7.01 eV, respectively; in qualitative agreement with the values 2.70, 2.72, 2.42, and 7.81 eV reported in Ref.~\cite{stoneham}.

The calculated formation energy of $V_{\rm Li}^{-}$ and $V_{\rm H}^{+}$ is 0.86 eV at $\mu_e^{\rm int}$, independent of the atomic chemical potentials. This corresponds to the well-known situation in defect chemistry, characterized by dominating Schottky disorder, with a formation energy of the dissociated ($V_{\rm Li}^{-}$,$V_{\rm H}^{+}$) Schottky pair of 2$\times$0.86 eV, independent of the chemical potentials of the components. This energy is slightly higher than the preliminary value (0.69 eV) reported in Ref.~\cite{hoang_prb2012} based on calculations that used smaller (64-atom) supercells and omitted the finite-size supercell correction. For these dominant point defects, we also performed calculations using the HSE06 functional. Although the HSE06 band gap is larger than the GGA gap, the calculated HSE06 formation energy at $\mu_e^{\rm int}$ is very similar to the GGA value: 0.84 eV in HSE06 compared to 0.86 eV in GGA. The fact that the formation energy at $\mu_e^{\rm int}$ is not very sensitive to the functional was also recently observed in a study on MgH$_2$ \cite{roy13}.

Inside the bulk, in the absence of any contact with a surface or interface, only the (H$_i^{-}$,$V_{\rm H}^{+}$) and (Li$_i^{+}$,$V_{\rm Li}^{-}$) Frenkel pairs can be created directly, but as shown in Table~\ref{tab} their formation energy is very high. The dominant point defects, $V_{\rm Li}^{-}$ and $V_{\rm H}^{+}$ vacancies, can also be created at a surface or interface where mass conservation is not required. The ($V_{\rm Li}^{-}$,$V_{\rm H}^{+}$) Schottky pair has a binding energy of 0.58 eV, but that does not necessarily make it more favorable to form as a unit; as discussed in Ref.~\cite{walle:3851}, if the binding energy is smaller than the formation energy of the constituents, entropy favors the formation of individual defects~\cite{walle:3851}.

Our results show that $V_{\rm Li}^{-}$ and $V_{\rm H}^{+}$ can equally contribute to ionic conductivity. The activation energy for self-diffusion of $V_{\rm Li}^{-}$ or $V_{\rm H}^{+}$ is 1.37 eV, which is the formation energy plus the migration barrier (see Table~\ref{tab}). For comparison, we also list in Table~\ref{tab} the activation energies associated with other defects and complexes.

Experimental values for the ionic conductivity were reported by Varotsos and Mourikis~\cite{Varotsos} and Ikeya~\cite{Ikeya}. Their analyses in different temperature regimes allowed them to extract a migration barrier for the current-carrying defect, with a value $E_m$=0.53 eV \cite{Varotsos} or 0.54$\pm$0.02 eV \cite{Ikeya}. This is in very good agreement with our calculated migration barrier of 0.51 eV for the $V_{\rm Li}^{-}$ and $V_{\rm H}^{+}$ defects (see Table~\ref{tab}). The experimental activation energy in the ``intrinsic'' region (at high temperatures) was found to be 1.695$\pm$0.005 eV \cite{Varotsos} or 1.70$\pm$0.10 eV \cite{Ikeya}. This is somewhat higher than our calculated value (1.37 eV) and could be indicative of the presence of an additional kinetic barrier.

The low formation energy of $V_{\rm H}^{+}$ (and $V_{\rm Li}^{-}$) also explains the hydrogen loss at the LiH surfaces and hydrogen emission as observed in experiments~\cite{siefken:3394,Powell1974345}. The processes can be understood in terms of the vacancies being injected into the bulk where their simultaneous presence ensures charge neutrality. Note that injection of $V_{\rm H}^{+}$ ($V_{\rm Li}^{-}$) into the material is equivalent to formation of H$^-$ (Li$^+$) on the surface. Hydrogen loss can occur in two ways. As addressed in Ref.~\cite{Powell1974345} (and also to some extent in Ref.~\cite{siefken:3394}), hydrogen can leave the sample in the form of H$_2$. This requires removal of an electron from the H$^-$ ions at the surface; this electron can be added to the Li$^+$ on the surface, which can then aggregate and form Li metal. In the experiment of Ref.~\cite{siefken:3394}, hydrogen is ejected in the form of H$^-$ ions by application of an extraction voltage. In this case the electron is thus ejected along with the hydrogen. This leaves Li$^+$ ions on the surface. The authors of Ref.~\cite{siefken:3394} explicitly state that in their experiment the sample is connected to a power supply which delivers negative charge, which neutralizes the Li$^+$. Therefore, in both cases charge neutrality is obeyed.

With regard to the amide/imide reaction, Eq.~(\ref{eq:amide}), the ease of $V_{\rm H}^{+}$ formation implies that LiH can provide H$^{-}$ ions at the LiNH$_{2}$/LiH interface. These H$^{-}$ ions then combine with H$^{+}$ ions from LiNH$_{2}$ to form H$_{2}$~\cite{hoang_prb2012}. Finally, in the reaction with the metal or intermetallic compound $M$ in Eq.~(\ref{eq:li-ion}), LiH can also provide H$^{-}$ ions at the $M$/LiH interface to form $M$H$_{x}$.

\section{Summary}

We have carried out a first-principles study of native defects in LiH. The negatively charged lithium vacancy and positively charged hydrogen vacancy are found to the dominant defects. The relatively low formation energy of these defects implies that LiH can easily provide Li$^{+}$ and/or H$^{-}$ ions for reactions between LiH and another reactant. Ionic conduction in LiH occurs via a vacancy mechanism with an activation energy of about 1.37 eV. Our results also explain the hydrogen surface loss phenomenon as observed in experiments.

\begin{acknowledgments}

This work was supported by the U.S.~Department of Energy (Grant Nos.~DE-FG02-07ER46434 and DE-FG52-08NA28921), by the U.S.~Naval Research Laboratory (Grant No.~NRL-N00173-08-G001), and by the Center for Computationally Assisted Science and Technology (CCAST) at North Dakota State University. Computing resources were provided by the Center for Scientific Computing at the CNSI and MRL: an NSF MRSEC (DMR-1121053) and NSF CNS-0960316, the National Energy Research Scientific Computing Center, which is supported by the DOE Office of Science under Contract No.~DE-AC02-05CH11231,
the Extreme Science and Engineering Discovery Environment (XSEDE), supported by NSF (OCI-1053575 and NSF DMR07-0072N), and by CCAST.

\end{acknowledgments}

\end{document}